# Boltzmann Brains—I'd Rather See Than Be One

J. Richard Gott, III

Department of Astrophysical Sciences, Princeton University, Princeton, NJ 08544
E-mail: jrg@astro.princeton.edu

**ABSTRACT**

A perceived problem with the standard flat-lambda model is that in the far future spacetime becomes an exponentially expanding de Sitter space, filled with Gibbons-Hawking thermal radiation, and given infinite time there will appear an infinite number of Boltzmann Brains (BB's) per finite co-moving volume today. If BB's outnumber ordinary observers by an infinite factor, why am I not one? This Gibbons-Hawking thermal radiation is observer dependent—due to observer dependent event horizons. Different observers moving relative to each other will see different photons, and different BB's. I will argue that the only particles that are real are the particles dredged out of the quantum vacuum state by particular real material detectors. (In much the same way, accelerated detectors dredge thermal Unruh radiation out of the Minkowski vacuum due to their observer dependent event horizons.) Thus, I may see a thermal BB, but cannot be one. Observer independent BB's can be created by quantum tunneling events, but the rate at which ordinary observers are being added to the universe by tunneling events to inflating regions exceeds the rate for producing BB's by tunneling by an infinite factor. I also argue that BB's do not really pass the Turing test for intelligent observers. Thus, the standard flat-lambda model is safe.

## 1 INTRODUCTION

The invasion of the Boltzmann Brains has become a perceived problem for the standard flat-lambda model. In the far future the standard flat-lambda model becomes dominated by the lambda term and approaches an exponentially expanding de Sitter geometry:

$$ds^2 = -d\tau^2 + r_o^2 \cosh^2(\tau)(d\chi^2 + \sin^2(\chi)[d\theta^2 + \sin^2\theta d\phi^2]) \tag{1}$$

where $r_o = (3/\Lambda)^{1/2}$.

This de Sitter geometry can be partially covered by Schwarzschild coordinates centered on a particular geodesic observer:

$$ds^2 = -(1 - r^2/r_o^2) dt^2 + (1 - r^2/r_o^2)^{-1} dr^2 + r^2(d\theta^2 + \sin^2\theta \, d\phi^2) \tag{2}$$

There is an event horizon at $r = r_o$, and because of this an observer at $r = 0$ will observe Gibbons and Hawking (1978) radiation at a temperature of $T = 1/2\pi r_o$. The Hadamard function tells us that a material detector located at rest at $r = 0$ will come into thermal equilibrium and attain an equilibrium temperature of $T = 1/2\pi r_o$. If it is an atom its energy levels will become populated with a Boltzmann distribution at that temperature. An observer at the origin will observe isotropic thermal radiation with a temperature of $T = 1/2\pi r_o$ at rest with respect to himself. From a thermal distribution one expects occasionally thermal fluctuations to produce Boltzmann Brains (as complex as a human brain). While these are exponentially rare, if one waits long enough one will eventually see one (since the de Sitter state lasts forever). Various authors have estimated how long one would have to wait on average to see one: $T_{av} \sim \exp(10^{50})$ (cf. Linde 2007). Dividing by the volume within the event horizon (the volume out to $r = r_o$) which is $8\pi^2 r_o^3$ gives the probability per unit four-volume of producing a Boltzmann Brain: $dP/dV_4 = 1/(8\pi^2 r_o^3 T_{av})$ which is finite. Consider a finite co-moving volume of our expanding universe today [say $(1\ h^{-1}\ \text{Gpc})^3$] which will harbor by definition a finite number of ordinary intelligent observers like ourselves. In the infinite future this region will expand by an infinite factor as the lambda term comes to dominate and ultimately it will achieve an infinite four-volume in the infinite future. Therefore it should harbor by the above argument an infinite number of BB's. Therefore BB's should outnumber ordinary observers like us by an infinite factor. So why am I not a BB?

A number of authors have proposed different solutions to this problem, as we shall discuss. The Boltzmann Brain problem is sufficiently serious that it has prompted Hartle and Srednicki (2007) to question whether we are typical, but all the other authors considering this question have implicitly assumed the Copernican Principle—or the Principle of Mediocrity as Vilenkin has sometimes called it. As laid out by Gott (1993) this runs as follows: the Copernican Principle states that it is unlikely for your location in the universe to be special *among* intelligent observers. Why? Because out of all the places for intelligent observers to be there are by definition only a few special places and many non-special places, so you are simply likely to be at one of the many non-special places. I defined an intelligent observer as one who is self-conscious, able to reason abstractly create art, etc., and ask questions like "How long will my species last?", or equally "Where am I in the Universe?" or "Am I an ordinary observer or a Boltzmann Brain?"

It is worth noting that the Copernican Principle has been one of the most successful scientific hypotheses in the history of science. When Hubble discovered that other galaxies were fleeing from us in an isotropic fashion, it could have been because our galaxy was in the center of a finite spherical explosion, but after Copernicus we were not going to fall for that—we couldn't be at a special place at the center of the universe. No, if it looked like that to us it must look that way from every galaxy—our location should not be special. That led to the homogeneous, isotropic big bang models which led to Gamow, Herman, and Alpher's prediction of the cosmic microwave background radiation which was later discovered by Penzias and Wilson. It was one of the greatest predictions ever to be verified in the history of science. All because of taking very seriously the idea that our location should not be special (cf. Gott 2001 for more discussion). We use the Copernican Principle all the time. Today, whenever we examine the cosmic microwave background we assume we are looking from a random location in space. Indeed, in examining the results from a scientific experiment we routinely assume that we are looking at a typical sample and not a particularly "lucky" sample.

Page, in a talk in Cambridge England in 2001, favored the idea of regarding one's current conscious perception as chosen randomly from all conscious perceptions. (In that case I then might have to wonder why I am not an ant. There are $10^{15}$ of them on Earth, and they outweigh human beings, and so even though their brains are small might have a comparable number of perceptions in total.) I would argue instead that one should regard my current *intelligent* observation as drawn randomly from the *intelligent* observations made by *intelligent* observers—observers able to formulate such questions as "am I an ordinary observer or a BB?" I *must* be an intelligent observer to participate in this discussion and so, by the Copernican Principle, my intelligent observation should not be special among such intelligent observations. That is the viewpoint I will take here: My intelligent observations are not special among those made by intelligent observers in the universe.

If BB's outnumber ordinary observers by an infinite factor, then if I am not special then I should likely be a BB. (Mind you most BB's make only one observation before expiring—but if they outnumber ordinary observers by an infinite factor that does not matter, their observations would also dominate the total, making my current observation special as well.) Since I am not a BB, something must be wrong. (I would note that the great majority authors considering this problem have been in agreement that the Copernican Principle should be upheld here and that we should not be among a finite number of ordinary intelligent observers in our co-moving volume if an infinite number of BB's brains occur later. In particular, no one considering the problem has argued that the BB's do not count because they do not exist yet, for example--a point that would be irrelevant in the multiverse in any case, where many de Sitter phases like the present one have preceded our own.) With the exception of Hartle and Srednicki, the many authors considering this problem have taken the spacetime view that we should be typical among intelligent observers, wherever or whenever they occur. I will take that point of view here.

One solution to this problem (cf. Dyson, Kleban & Suskind [2002]) has been proposed by Page (2006). It is the rather dramatic conclusion that the cosmological model must therefore be wrong. The de Sitter phase must not last forever. The de Sitter phase is unstable via tunneling to the formation of bubble universes of lower vacuum density. Within this infinitely expanding inflationary sea, an infinite number of bubble universes of lower density vacuum state can form (cf. Gott 1982). Each bubble wall accelerates outward with constant acceleration (because the density within the bubble is less than the density outside and the pressure is less negative than outside, so the extra negative pressure from outside pulls the bubble wall outward), giving it the shape of a hyperboloid of one sheet in spacetime. At late times the bubble wall expands at nearly the speed of light, but the forming bubbles fail to percolate and fill the space if the bubble formation rate per four-volume is less than a critical value [which must be less than 0.24 $r_o^{-4}$ ](cf. Guth and Weinberg 1982, Gott and Sattler 1984). Dark energy we suspect is in the form of an effective cosmological constant, a vacuum energy density of V($\phi$) sitting at a local minimum giving an effective lambda term today, where ϕ is a scalar field for example. This may not be the global minimum. In fact, there may be many minima, with values of V that are smaller than the current one. These may be either positive (leading to other de Sitter bubbles) or zero (leading to bubbles with Minkowski vacuum inside) or negative (leading to bubbles with anti-de Sitter vacuum inside). Conventional wisdom says the negative density anti-de Sitter bubbles are sterile—since they resemble open cosmologies that although initially expanding, recollapse after a short finite time. If these other lower density vacuums exist, after long enough time the solution will tunnel to one of them and a bubble of lower density vacuum will form. The formation rate of lower density vacuum density bubbles per unit four-volume $r_o^4$ is expected to be exponentially small due to the exponentially small barrier penetration probability. In this situation the bubbles do not percolate, and the de Sitter phase continues forever, leading to the BB situation. Page proposes that, instead, the bubble formation probability is higher than the critical value for percolation, that is, of order $1/r_o^4$ or greater. That means that the de Sitter phase will end in about 20 billion years. Page specifies tunneling to nothing, with bubbles of nothing expanding to percolate and end the de Sitter phase in 20 billion years. If the de Sitter phase is not ended by then (if the bubble formation rate is lower) then the bubbles do not percolate and there is no stopping the de Sitter phase expanding forever and the BB situation arises. To achieve Page's result, the bubbles would either have to be filled with nothing or sterile universes and the barrier penetration probability would have to be surprisingly small.

Another possibility, perhaps more easily realized would be to simply adopt the cyclic model of Steinhardt & Turok (2002) a point that has been independently noted by Carlip (2007). In the cyclic model the current de Sitter phase ends with reheating and the formation of a new big bang cycle after only a finite period of say a trillion years. That gives a finite four volume for the de Sitter future of our piece of co-moving volume, of say $(1\ h^{-1}\ Gpc)^3$, and given the exponentially small probability of finding a BB in that finite region one would not expect on average for any BB's to appear in this region. Therefore the BB problem is solved. Is the fact that I am not a BB evidence favoring the Steinhardt & Turok model? Alternatively, there may instead be a slow variation of the fundamental constants (Carlip [2007]), or a slow fall off in V(ϕ) as it rolls down a hill to zero, ending the de Sitter expansion.

Still another possible solution to the BB problem (discussed by Linde [2007]) would be to note that in inflation new bubble universes are forming all the time in an inflating sea. If you slice spacetime along a hypersurface of fixed proper time, you will find, at the epoch we are at, exponentially more young bubble universes--filled with ordinary observers--than old ones already in the de Sitter phase filled with BB's. Since the inflationary timescale is so short—of order $10^{-34}$ sec, universes that are younger by a time Δt, are more common by a factor of $\exp(3\Delta t/10^{-34}\ sec)$ than older ones. Since $10^{-34}$ sec is so short, the many newly formed ordinary observers dominate over the few BB's at a given epoch and the BB problem would be solved, since in eternal inflation each and every epoch is identical. A problem with this solution is the youngness problem. In other words I should find myself very early in my universe. It would make it likely for me to be the first member of my intelligent species which itself should be formed very early in the history of the universe, since universes in which intelligent species were first forming should vastly outnumber [by a factor of exp($3\Delta t/10^{-34}$ sec)] older ones where intelligent species were already formed at some time $\Delta t$ in the past. I find myself formed some 200,000 years after the formation of my species and some 13.7 billion years after the big bang—so this formulation must be incorrect. Thus, Linde argues that instead of counting probabilities one should count probability fluxes.

Measures relying on a single time slice in the multiverse have been criticized by Vilenkin (1995) and Aguirre, Gratton & Johnson (2006) on the grounds that the results depend on the time slice. If slices of constant proper time are taken, one reaches after a while, one stationary solution, while if one uses slices of constant Hubble expansion time, one gets after a while another different stationary solution with different relative probabilities for formation of different types of pocket universes.

Vilenkin (2006) instead proposes that one should compare the rate at which equilibrium de Sitter BB fluctuations occur per unit four-volume in de Sitter space with the rate per unit four-volume that new inflating universes are formed--multiplied by the number of observers formed in each inflating universe by non-equilibrium processes later. Garriga and Vilenkin (1998) have noted that bubbles of higher than the de Sitter density can form by tunneling. Such a bubble has a higher vacuum density inside and a more negative pressure inside than outside, and so begins to collapse, but if its initial radius is larger than $r_o$ even as the collapsing bubble wall approaches the speed of light, the wall is not able to complete the collapse because the outside space is expanding so fast. The collapsing bubble reaches an asymptotically fixed co-moving size in the outside space and thus grows to infinite size in the infinite future. Since each such inflating bubble will create a new region of inflation—a new inflating sea—which will expand indefinitely, it will create an infinite number of ordinary pocket universes like ours, and an infinite number of ordinary intelligent observers. Aguire, Gratton, & Johnson (2006) have noted that another process for forming an inflating region is quantum tunneling to a Schwarzschild—de Sitter geometry with an inflating region on the other side of an Einstein-Rosen bridge (the geometry discussed by Farhi, Guven, & Guth [1987] for creating a universe in the lab). The probability for forming this wormhole is exponentially small but the inflating region can expand forever and give rise to an infinite number of pocket universes and an infinite number of ordinary observers. The probability of forming a BB in de Sitter space is exponentially small, and the probability of forming a new inflationary region (at the grand unified density) is also exponentially small but the inflating region produces an infinite number of ordinary observers by non-equilibrium processes later and thus the number of ordinary observers outnumbers the number of BB's by an infinite factor. I am therefore likely to be an ordinary observer. The fact that there is a time delay in forming the ordinary observers whereas the BB forms at once, is not important Vilenkin argues. On the other hand, the new inflating region will also create new de Sitter regions in which an infinite number of BB's are also created—and new inflating regions ad infinitum. So if delay is not important then where does one stop? Vilenkin calls a stop by *only* counting the number of observers created by *non-equilibrium* processes later. Thus the equilibrium fluctuation producing a BB gets a weight of 1, while the tunneling event giving rise to inflation gets an infinite weight because it produces an infinite number of ordinary observers by non-equilibrium processes later. The BB's produced later (in de Sitter phases) by the new inflationary region don't count because they are *not* formed by non-equilibrium processes. Of course, that might give the original BB a weight of 0 instead of 1, because it is not formed from a non-equilibrium process either. This prescription seems unfair to the BB's. I am an observer formed by non-equilibrium processes, so it seems to unfairly count only observers (like me) formed later by non-equilibrium processes while not counting thermal BB's formed by equilibrium processes later. If BB's can be formed by equilibrium processes later due to the formation of the new inflating region, then why shouldn't they also count?

Inflation also produces a de Sitter geometry and a spacetime filled with much hotter Gibbons & Hawking thermal radiation, so why are BB's not a problem in the inflationary stage that proceeds our own? The reason is that the e-folding time is small ($10^{-34}$ sec) so the event horizon has a circumferential radius of only $r_o \sim 3 \times 10^{-24}$ cm. This is too small to make a causally connected intelligent observer. Bousso & Freivogel (2006) note that BB's are most frequently produced in string landscape vacuua where $r_o \sim 1$ m, just large enough to contain a reasonably complex BB. The average time to encounter a BB in such a vacuum is $T_{av} \sim \exp(10^{45})$ they argue.

Bousso and Freivogel (2007) have a different approach to the counting problem. They argue that one should take a more local approach and examine what one eternal observer would observe. Just as the quantum state is defined outside a black hole, they argue the quantum state can be defined inside the event horizon in de Sitter space. One eternal observer, traveling on a geodesic (i.e. sitting at rest at r = 0) will see the static Schwarzschild metric in equation 2, out to the event horizon. The proper question Bousso and Freivogel argue is how many BB's would that eternal observer see, given an infinite amount of time. If the probability for formation of a new inflating bubble per unit four-volume is larger than the probability of forming a BB, the eternal observer will typically see the de Sitter phase end, and will pass into the new region before encountering any BB's at all. Earlier he would have encountered a finite number of ordinary observers, so they would by his counting outnumber BB's. The probability of forming a new inflating bubble and a BB are both exponentially small and it is not clear which process would win, but this at least offers a hope of solving the BB problem. (One difficulty with the eternal observer approach as pointed out by Aguirre et al. 2006 is that eternal observers will miss seeing new inflating regions formed by the Farhi, Guth, and Guven, creation of a universe in a lab mechanism, and thus will unfairly not count them when evaluating the frequency of different types of vacuum regions.) But by focusing on how many BB's are seen, Bousso and Freivogel (2007) have taken an important new tack, a viewpoint I will find valuable in one way as indicated below. However, one may note, that while the space visible to a single eternal observer remains of fixed volume per unit time (a spacetime cylinder with radius of the event horizon) the actual co-moving volume beyond this is expanding exponentially with time, so many BB's are hiding beyond the event horizon. And the question is why am I an ordinary observer rather than a BB? Not whether I could see a BB, but whether I could *be* one. As the material particles in our co-moving volume thin out exponentially, they will only be able to see an infinitesimal fraction of the whole future co-moving volume (within their individual event horizons). Almost all of the BB's would remain hidden behind these event horizons, like birds in a rainforest unseen by birdwatchers. Do they count? Could I, as a random intelligent observer be a BB alone and unseen in de Sitter space? Remember, the intelligent observer only has to be self-conscious—and is not required to be seen by others.

**2 UNRUH RADIATION**

Suppose instead of finding ourselves in a universe dominated by dark energy, we found ourselves in an open universe with zero cosmological constant and $\Omega_m < 1$. Then the universe would expand forever and at late times would approximate a Milne cosmology with metric:

$$ds^2 = -d\tau^2 + a^2(\tau)[d\chi^2 + \sinh^2\chi(d\theta^2 + \sin^2\theta\, d\phi^2)] \qquad (3)$$

where $a(\tau) = \tau$ (the expansion becomes linear as the universe approaches zero density)

Setting $t = \tau \cosh\chi$, $z = \tau \sinh\chi \cos\theta$, $y = \tau \sinh\chi \sin\theta\cos\phi$, $x = \tau \sinh\chi \sin\theta\sin\phi$ (4)

we find: $ds^2 = -dt^2 + dx^2 + dy^2 + dz^2$ \qquad (5)

which is a Minkowski metric, as expected, for it is an asymptotically flat spacetime with zero density and pressure. This is a spacetime which has a temperature of $T = 0$ in the limit as $\tau \to \infty$. If this were the cosmology we were living in there would be no worry over BB's. With a temperature of zero, in this spacetime as $\tau \to \infty$, no one would be worrying about there being any BB's. And yet we can find them if we go looking. The origin of the Minkowski coordinates can be shifted to some late event in this space time. Then establish new Rindler (1966) coordinates $(T, \sigma, x, y)$:

where: $t = \sigma\sinh(T/\sigma_o)$, $z = \sigma\cosh(T/\sigma_o)$ where $\sigma_o$ is a constant \qquad (6)

Expressed in Rindler coordinates the Minkowski metric becomes:

$$ds^2 = -(\sigma/\sigma_o)^2\, dT^2 + d\sigma^2 + dx^2 + dy^2 \qquad (7)$$

In Rindler coordinates, the metric is static (the coefficients of the metric depend only on the radial coordinate $\sigma$ which measures proper distance from the origin). An atom on a worldline with $\sigma$ = const., is accelerated with acceleration $a = 1/\sigma$. Its worldline is a hyperbola $z^2 - t^2 = \sigma^2$. It maintains a constant spacelike separation of $\sigma$ from the origin. An atom on this hyperbolic worldline will receive no light signals from events with $t > z$. There is an observer dependent event horizon for this atom (the null surface $t = z$). The Hadamard function shows the atom traveling on this accelerated world line will see thermal Unruh (1976) radiation with a temperature $T_U = 1/2\pi\sigma = a/2\pi$. The renormalized stress energy tensor is $\langle T_{\mu\nu}\rangle_{ren} = 0$, because we have the Minkowski vacuum state. However the atom on the accelerated trajectory will see a Rindler vacuum state plus Unruh thermal radiation: $\langle T_{\mu\nu}\rangle_{ren} = 0 = \langle T_{\mu\nu}\rangle_{thermal} + \langle T_{\mu\nu}\rangle_{Rindler\ vacuum}$. The Unruh thermal radiation has a pressure equal to 1/3 the energy density ($\rho_{th} \propto T_U^4 \propto 1/\sigma^4$) in each direction, and so the Rindler vacuum has an energy density which is negative and a negative pressure equal to 1/3 of that negative energy density. The accelerated observer (or atom) will see Unruh thermal radiation plus a vacuum polarization whose $T_{\mu\nu}$ is the negative of that of the Unruh thermal radiation making the sum zero. The accelerated observer will agree that the total renormalized stress energy tensor is $\langle T_{\mu\nu}\rangle_{ren} = 0$. He just sees thermal photons that have been dredged up out of the Minkowski vacuum to become real particles as far as he is concerned. If the accelerated atom is considered a simple detector with two energy levels the occupation numbers will settle on the thermal Boltzmann distribution $N_2/N_1 = \exp(-\Delta E_{12}/kT_U)$. The atom interacts with these thermal Unruh photons. By the equivalence principle, this is as if the atom is at rest in a static gravitational field (the Rindler metric is a static metric) immersed in a thermal bath. The thermal bath this accelerated observer sees at a temperature $T_U$ has the usual thermal fluctuations, and so very occasionally, he should see a BB. If he continues accelerating forever, the thermal state will last forever and he will eventually see BB's. Of course it would take an infinite amount of rocket fuel to keep him accelerating at constant acceleration forever.

But here is the point. If an accelerated observer along a particular hyperbolic worldline in Minkowski space would have detected a BB, would that BB still exist (and wonder why it was a BB) if that accelerated observer (or atom) did not exist. I would argue that that is not the case. The photons absorbed by an accelerated atom in Minkowski space which cause one of its electrons to jump from one energy level to another are dredged out of the Minkowski vacuum state by the accelerated motion of that atom. These are real photons produced by the Unruh process and absorbed by the atom. But absent that accelerated atom, the supposed division between Unruh radiation and Rindler vacuum at a given event is observer dependent. The event horizon producing the Unruh radiation is observer dependent. It depends on that accelerated observer (or atom in this case). A different accelerated observer passing through the same event, whose acceleration was in a different direction and of a different magnitude, would have a different event horizon and would see different Unruh photons. A more highly accelerated observer would see more and hotter Unruh photons. Absent that accelerated atom, those particular Unruh photons are not real.

In special relativity, the time ordering of two events with a spacelike separation is observer dependent, and there is no real answer as to which occurred first. Real things, like the timelike or spacelike separation between two events as measured by the metric, are those which are observer independent. So we should not think of all the possible BB's in Minkowski space that could possibly be observed by all possible hypothetical accelerated observers as real. What is real are the Unruh photons that are actually detected by a particular real accelerated detector made of normal atoms. All observers looking at that detector will see the detection. That is observer independent.

# 3 OBSERVER DEPENDENT GIBBONS & HAWKING RADIATION IN DE SITTER SPACE

The renormalized energy momentum tensor in de Sitter space for the Gibbons & Hawking vacuum state is proportional to a small cosmological constant. For the case of a simple scalar field (cf. Bunch & Davies [1978], Page [1982], Bernard & Filacci [1986], Gott & Li [1998]):

$$<T_{\mu\nu}>_{ren} = -g_{\mu\nu}(1/960\pi^2 r_o^4) \tag{8}$$

The Hadamard function shows that a geodesic detector will be in thermal equilibrium with a thermal bath at a temperature of $T = 1/2\pi r_o$ (Gibbons & Hawking [1978], Birrell & Davies [1982]). With photons and gravitons and other particles included, the absolute magnitude of this cosmological constant would be changed but the renormalized energy momentum due to these fields would still be proportional to a cosmological constant because a cosmological constant is the only thing that is de Sitter invariant. There can be no preferred state of rest for the renormalized stress energy tensor, because the de Sitter solution has no preferred state of rest. So we expect

$$8\pi <T_{\mu\nu}>_{ren} = -g_{\mu\nu} C_1 r_o^{-4} - g_{\mu\nu} \Lambda \tag{9}$$

Where $\Lambda$ is the value of the cosmological constant produced by the $V(\phi)$ of the dark energy. And $C_1$ is a constant of order unity. This gives an effective cosmological constant

$$\Lambda_{eff} = \Lambda + C_1 r_o^{-4} \tag{10}$$

Now to solve Einstein's field equations we require $r_o = (3/\Lambda_{eff})^{1/2}$ so:

$$r_o^{-2} = [3 \pm (9 - 4C_1\Lambda)^{1/2}]/2C_1 \tag{11}$$

where we are interested in the smaller solution where

$$r_o^{-2} = (\Lambda_{eff}/3) \approx (\Lambda/3)[1 + (C_1\Lambda/9)] \tag{12}$$

(cf. Gott and Li 1998 for discussion). The Gibbons & Hawking vacuum state makes a slight correction to the cosmological constant $\Lambda$ due to a scalar potential ($V(\phi)$). A geodesic observer stationary at $r = 0$ will see thermal radiation at a temperature $T = 1/2\pi r_o$ at rest with respect to himself. These thermal particles (primarily photons and gravitons of wavelength of order $r_o$) will have a stress energy tensor $<T_{\mu\nu}> =$ diag$(1,1/3,1/3,1/3)C_2 r_o^{-4} \propto$ diag$(1,1/3,1/3,1/3)T^4$ where $C_2$ is a constant of order unity. The energy density in these photons is tiny. The total stress energy tensor seen by the geodesic observer may be broken up into three parts: a cosmological constant term [$(-g_{\mu\nu}\Lambda)/8\pi$], thermal radiation [diag $(1,1/3,1/3,1/3)C_2 r_o^{-4}$], and vacuum polarization [$-g_{\mu\nu} C_1 r_o^{-4}/8\pi$ - diag $(1,1/3,1/3,1/3)C_2 r_o^{-4}$]. Together they add up to [$-g_{\mu\nu}(\Lambda + C_1 r_o^{-4})]/8\pi$ which is proportional to a cosmological constant and is de Sitter invariant. All observers agree on this total stress energy tensor which enters the Einstein equations. Our geodesic observer at rest at $r = 0$ will see thermal radiation that is at rest with respect to him, and a vacuum polarization part that looks like a small cosmological constant minus that thermal radiation. Another geodesic observer passing through the same event as our geodesic observer with a velocity v with respect to him will observe thermal radiation at rest with respect to her and will make a different division of the stress energy tensor into thermal radiation and vacuum polarization parts. Our first observer will see different photons than seen by our second observer. If the original observer were to find a BB the second observer would not see that same detailed distribution of particles. The two observers see different thermal photons. The thermal photons are observer dependent because each observer has a different observer dependent event horizon (one that surrounds their world line with a cylinder of circumference $2\pi r_o$). If an observer finds (rarely) a BB in the thermal radiation he sees, it is quite likely to be nearly at rest with respect to him. Finding a BB in that thermal distribution with a high γ factor would be even less likely, since it would require even more energy to produce. Another observer traveling at high γ factor with respect to the first will observe a thermal distribution of photons at rest with respect to her, and not see the same photons seen by the first observer. The BB has a complicated brain whose intricate details depend on incredible good luck in the photons and particles seen with small probability in that thermal radiation (there will be a small probability in that thermal radiation for having some baryons etc. and then ultimately a BB). The second observer flying by at high γ factor will not see those same thermal radiation particles and will not agree that that BB is there. The BB's that are seen (rarely) by individual observers are observer dependent and therefore not real. What is real are the photons or particles that are actually detected by the first observer. Any photons he detects, he has dredged out of the quantum vacuum and registered on his detector and the second observer will see those changes in his detector.

Suppose the first observer sees a BB waving at him and takes a picture of it. The second observer will not see that BB but will see the photograph of it the first observer has taken. Now consider the first observer. He has seen a BB waving at him. This could be produced by a BB materializing out of that thermal bath and then waving at him, sending photons with that view on their way to him. That would be a relatively rare event as expected. But less rare, would be just those photons materializing out of that photon bath headed toward him with that pattern. They do not have as much energy as the BB and have less information in them than the BB itself would have. The Copernican Principle tells the observer that he is not likely to be special among observers who have seen BB's waving at them. Of all those observers, most have simply seen photons headed toward them that look like a BB. So if you see a BB from a thermal bath you should conclude that most likely it is just photons coming toward you that look like a BB rather than a BB itself (which would be very much more unlikely). But it is even worse than that. The photons and other particles that are not detected are observer dependent and would not be seen by other observers. So there would not be agreement that they were real.

In the everyday world, when we see a person waving it is most likely because there is a person waving at us, rather than just photons in a lucky configuration headed toward us looking like that. If we pick up a book in the library and read the first half of it and it appears to be Hamlet, then it is quite likely that the second half of the book is the second half of Hamlet. But if we are told that all the books in the library have been typed randomly by monkeys, (which is the analogy with BB's) we would expect that the second half of the book was most likely nonsense. Likewise, in the BB case, extrapolating beyond the information we have actually detected, to more information beyond is not usually justified.

So, Bousso and Freivogel seem to be on the right track in one sense. It is important what is seen by real observers. So even when a BB is seen, it is only the detected photons that are real and not the BB itself (which is observer dependent). Even if an observer sees a BB, it does not mean that one is there (thinking its own thoughts).

Temperature in de Sitter space should be defined locally by observers and not globally (c.f. Narnhofer, Peter & Thirring [1996]) because the observed temperature depends on the acceleration of the detector. For example, consider an accelerated observer who is stationary in the coordinate system of equation (1) at a radius r = const. > 0 where $r < r_o$. Such an observer will see a thermal bath (at rest with respect to himself) with a temperature of $T = 1/[2\pi r_o(1 - r^2/r_o^2)^{1/2}]$ (c.f. Gott 1982). If one constructs a rigid rod of proper length $L = 2r_o \arcsin(r/r_o)$ centered on the origin at r = 0, its ends will be accelerated by the force applied by the rod so as to stay stationary at radius r from the center in Schwarzschild coordinates (c.f. equation 2). Radiation traveling from the static location r = r to the location r = 0 will redshift by the proper amount according to the variation of $g_{oo}$ expected in a static gravitational field so that arriving at r = 0 it will have the proper temperature $T = 1/2\pi r_o$. As $r \to r_o$, stationary (accelerated) observers at r = const. see $T \to \infty$, and we approach the flat spacetime situation where a highly accelerated observer has a hyperbolic worldline and sees Unruh radiation. Such an accelerated observer will see hotter radiation, and will have a larger chance of seeing a BB's. But again this is observer dependent radiation; different observers with different velocities and different accelerations would see different photons. Different observers will see different BB's and they would disagree about whether a BB was present at a particular location. The only things that are real are observer independent-- things that everyone can agree on.

[The black hole case is different. Hawking radiation which appears far from the hole is observer independent because it arises from an observer independent event horizon (the boundary of the past of future null infinity). The Hawking photons at large distance from the hole carry energy away from it causing its mass to slowly evaporate. This will occur whether or not those photons are ever detected. Those photons at large distances from the hole are real (observer independent).]

In the case of de Sitter space, the thermal radiation seen by various real observers is observer dependent and the only photons that are real are those actually detected by those real observers—those are observer independent and leave a record. Those observers may (occasionally) see a thermal BB, but the BB is not observer independent (therefore not real) and is not an intelligent observer thinking about the encounter.

**4 THE TURING TEST**

Are we are violating the Turing test here? Turning proposed that if one had a conversation with an entity [by typing questions], one should regard that entity as an intelligent observer if one could not distinguish it from a human being. This is an *external* behavioral test. So if an observer sees a BB and it passes the Turing Test, shouldn't we regard it as an intelligent observer? If the computer HAL in 2001 could, by repeated questioning prove indistinguishable from a human being, Turing would say HAL should be regarded as an intelligent observer—and subject to the same protections and rights as a human being. Our observer might have an extended conversation with the BB he sees, asking it say 20 questions which would be answered appropriately—just as a human would do. There is some tiny chance of finding that. (The answers to his questions, which he can record on his computer, are observer independent and therefore real.) But what happens if he asks a $21^{st}$ question? According to the theory of BB's, most BB's last only an instant before vanishing. They have very short lifetimes. Exponentially rarely, one will find one that can answer 20 questions in a row as a human would do, without vanishing or producing nonsense answers. This is, by definition, a very special long lived BB. It is unusual among BB's. You have to be very lucky (or equivalently wait a very long time) to see one. But BB's that will answer 21 questions in a row are exponentially rarer than ones that can answer 20 questions in a row. Or more precisely, getting 21 questions in a row correct from a thermal bath is exponentially rarer than having it answer 20 questions in a row correctly. So by the Copernican principle, if you are not special among those seeing a BB having just answered 20 questions in a row satisfactorily, it is quite likely that you will see the BB failing to answer the next question (the $21^{st}$) successfully. (Most BB's able to answer 20 questions can't answer the $21^{st}$.) No matter how many questions are answered successfully, the BB you see is likely to fail to answer the next one successfully (either by vanishing or by answering in nonsense). The BB you see does not pass the Turing test, a test which allows you to continue asking questions.

Indeed it is how I know I am not a BB. It might be argued, that if BB's outnumber ordinary observers by an infinite factor, that I should be likely to be a BB. In fact, it might be argued that I am in this case actually a BB. In other words, in this picture, all the intelligent memories that I have accumulated up to now, should have been exactly accumulated by an infinite number of BB's emerging from a finite co-moving volume in our universe in the future. Then it could be argued, that for every ordinary me there are an infinite number of BB's that have exactly the same mental processes and memories and experiences that I have had up to this moment. Therefore, if I am not special, I should be one of that infinite number of BB's. So how do I know that this is not true? How do I know that I am an ordinary observer, rather than just a BB with the same experiences up to now? Here is how: I will wait 10 seconds and see if I am still here. 1,

2, 3, 4, 5, 6, 7, 8, 9, 10 …. Yes I am still here. If I were a random BB with all the perceptions I had had up to the point where I said "I will wait 10 seconds and see if I am still here," which the Copernican Principle would require—as I should not be special among those BB's—then I would not be answering that next question or lasting those 10 extra seconds. BB's do not pass the Turing test for intelligent observers, so if I see one I should not regard it as an intelligent observer.

**5 TUNNELING**

There is another way to make Boltzmann Brains in de Sitter space. It is via quantum tunneling. Across a spacelike hypersurface the solution tunnels directly to a new Schwarzschild de Sitter geometry which has a Boltzmann Brain mass sitting in the de Sitter geometry. We can calculate the tunneling probability per unit four-volume knowing geometry of the two solutions. The tunneling probability is exponentially small. It can also be estimated by entropy arguments. The wait time for tunneling to produce a 10 kg BB is of order $T_{av} \sim \exp(4 \times 10^{69})$ (c. f. Bousso & Freivogel [2006]). This is a permanent change in the geometry of spacetime that remains after the tunneling (which has a Schwarzschild-de Sitter geometry corresponding to the 10 kg mass sitting in de Sitter space). The Boltzmann brain may disintegrate after a short time, but the mass making it up will remain afterward. (Making a BB out of the observed thermal radiation in de Sitter space is a fluctuation in the energy in the observed thermal radiation, but this may be in principle offset by a fluctuation lowering the energy level in the vacuum polarization so that the total stress energy is not changed. Even if the total energy does go up, consistent with the uncertainty principle, the fluctuation would go back down quickly and in any case different observers would not see the same detailed BB. As we have noted, such BB's are not observer independent.) The new BB created by tunneling, however, can be observer independent, for its mass changes the geometry of spacetime which everyone can agree on. This tunneling event adds one BB to the universe.

This tunneling event competes with other tunneling events which can also occur. Since a BB is roughly of order 10 kilograms in mass, the probability of forming one per unit four volume is similar to forming a 10 kilogram mass inflating "universe in a lab" by the Farhi, Guven, & Guth [1987] tunneling mechanism, if the inflating state is near the Planck density as would be implied by Linde's (1983) chaotic inflation model.
There is also an exponentially small probability for tunneling to an inflating bubble state that is at the GUT density. If this high density bubble is larger than the event horizon, even though the walls of this region are being pulled inward at the speed of light, it will inflate forever and create an infinite number of ordinary pocket universes like ours. In either type of tunneling event, if the inflationary state is not metastable, according to Linde's (1983) chaotic inflation, it forms an infinite number of pocket universes like ours by just rolling down hill without any other tunneling events. That produces an infinite number of ordinary observers. In favor of chaotic inflation is the fact that the WMAP results are consistent with a simple $V(\phi) \propto \phi^2$ chaotic inflationary potential (Spergel et al. 2007). If there is either tunneling to a "universe in a lab" inflating state, or to a bubble inflating state of radius greater than $r_o$ (which we expect to be less likely), an infinite number of ordinary observers will be created with no further tunneling events required.

(If the inflationary potential is metastable, the situation is more complicated. A tunneling event is then required to form a single bubble universe from the inflating state. So we are not comparing two tunneling events in a de Sitter space dominated by a cosmological constant—which is easy—but rather a tunneling event in an inflating region versus a tunneling event in a de Sitter space dominated by a small cosmological constant. Both the inflating region and the de Sitter region dominated by a cosmological constant have infinite four-volumes. One would then compare tunneling rates per unit four-volume. Both are exponentially small. One tunneling event could produce one observer independent BB, the other tunneling event would lead to formation of a bubble universe. A single bubble universe would be open and produce an infinite number of ordinary observers. But bubbles hit other bubbles, if the bubble in question joins a previously formed bubble cluster then it will add intelligent observers of its own, but perhaps subtract some that would have formed in other bubble universes had it not formed. On average, of course, intelligent observers are created by this process. A cluster containing an infinite number of bubble universes will produce an infinite number of ordinary intelligent observers. Whether the number per bubble is finite or infinite would have to be investigated. But it is large if the tunneling probability is exponentially small because a large co-moving volume is created in each bubble universe before others hit it. In any case, chaotic inflation seems to have the edge observationally with WMAP, so it appears that these complications needn't be considered.)

We are interested in the rate new ordinary intelligent observers versus observer independent BB's are being added to the universe by various tunneling events. We should do this by multiplying the tunneling probabilities per unit four-volume in de Sitter space by the change in ordinary intelligent observers (or observer independent BB's) added by each tunneling event. The tunneling probabilities for producing a BB and for producing an inflating region are both exponentially small, but the inflating region produces an infinite number of ordinary observers as opposed to the one observer independent BB. Therefore ordinary intelligent observers are being added to the universe by these tunneling events at a faster rate than observer independent BB's. Now this is a new version of Vilenkin's (2007) argument in full force. The qualms one had about it before were that the inflating region produces eventually an infinite number of thermal BB's also in later de Sitter phases, but now we know those BB's are observer dependent and don't count as intelligent observers. So the tunneling event forming an inflating region *changes* the future universe by adding intelligent observers in the form or ordinary observers, but adds no new intelligent observers that are BB's. *Future* additional tunneling events would be required to produce observer independent BB's. Here we are calculating the rate at which ordinary intelligent observers and observer independent BB's are being added by changes produced by the two types of tunneling events, so we should consider only the changes that follow from these tunneling events, supposing no future tunneling events were to occur. We then get the rate at which new ordinary intelligent observers and observer independent BB's are being added to the universe by those two types of tunneling events. The ratio of the tunneling probabilities is finite. An infinite number of ordinary intelligent observers are added by each inflationary tunneling event versus only one observer dependent BB for a BB tunneling event. Thus, the numbers of ordinary intelligent observers being added to the universe as a function of time by tunneling events is infinitely large relative to the number of observer dependent BB's added by tunneling events. (Thus, regardless of whether we regarded observer independent BB's formed by tunneling as intelligent observers, I should still not be surprised to be an ordinary intelligent observer.)

We should therefore expect ordinary intelligent observers to outnumber observer independent BB's in the universe by a large (infinite) factor. Mind you an observer independent BB, while real (because it is seen by all observers), still does not pass the Turing Test, and may not be regarded as an intelligent observer. However one may have a rarer tunneling event where instead of a BB one makes a huge mass with luckily just the initial conditions to make an entire solar system. In such a solar system (living in de Sitter space), intelligent life and intelligent observers may develop by ordinary evolutionary processes on a habitable planet. These intelligent observers would be able to answer a continuing sequence of questions and would pass the Turing Test and would qualify as intelligent observers. I would call them ordinary observers who happened to be born in a solar system that was not born in a standard inflationary big bang but one that by tunneling suddenly found itself in an empty de Sitter space. I am not born in such a solar system. And if I am not special then most solar systems harboring intelligent observers should originate in standard inflationary big bangs rather than quantum tunneling events. Recall that since a quantum tunneling event producing an inflationary region produces an infinite number of universes and solar systems made from standard inflationary big bangs, versus only one solar system made alone in de Sitter space from its tunneling event, the number of big bang solar systems would outnumber solar systems alone in de Sitter space by an infinite factor. I should not be surprised to live in a solar system born in a standard inflationary big bang rather than in an isolated quantum tunneling event. Inflation is just very efficient at producing solar systems and intelligent observers so it wins.

## 6 SUMMARY

The Gibbons & Hawking thermal radiation that is seen by geodesic observers in de Sitter space, and from which BB's are supposed to occasionally arise, is observer dependent. Two geodesic observers crossing paths with a relative velocity difference will see different Gibbons & Hawking photons. Each will see a distribution of photons at rest with respect to themselves. They have different observer dependent event horizons and see different photons. If one saw a thermal BB, the other would not be expected to observe it as well. Arguably the only things that are real are the photons actually detected by the real observers. If an observer takes a picture of a BB, which would occur occasionally, the picture would be real and the photons that camera dredged out of the quantum vacuum state would be real—and observer independent—but the BB itself would not. Just because you see a BB in de Sitter space, does not mean that *it* is thinking about the encounter  The BB does not count as an intelligent observer because it is observer dependent and does not pass the Turing test. It can be distinguished from a human because even if it answers 20 questions successfully in a row, it will likely fail to answer the next question. Real observer independent BB's can be created by tunneling events in de Sitter space. These change the geometry of spacetime and are observer independent. The rate at which new observer independent BB's are being added to the universe compared to ordinary intelligent observers can be estimated by comparing the (exponentially small) tunneling rate for creating a single BB to the (exponentially small) tunneling rate to an inflationary state multiplied by the number of ordinary observers thus created. With chaotic inflation, the one tunneling event to an inflationary state leads to an infinite number of ordinary observers (and no new observer independent BB's) with no further tunneling events. So the number of ordinary observers added by just that tunneling event is infinite. Thus the number of ordinary observers outnumbers the number of observer independent BB's being added by tunneling.

Therefore, the standard flat-lambda model, which has a de Sitter future, is not inconsistent with the Copernican principle and the fact that I am an ordinary observer. I might see a thermal BB but these are observer dependent so I could not *be* a de Sitter thermal BB. Observer independent BB's formed by tunneling are infinitely rare compared to ordinary observers produced by inflation. Even those observer independent BB's fail the Turing test, so it could be argued that as an intelligent observer I could not be one of them either. So with apologies to Gelett Burgess (1895), who wrote a similar poem about a Purple Cow, I could sum up the situation as follows:

I never saw a Boltzmann Brain;
I never hope to see one;
But I can tell you, anyhow,
I'd rather see than be one.

## REFERENCES

Aguirre, A., Gratton, S. & Johnson, M. C. (2006), "Hurdles for Recent Measures in Eternal Inflation, arXiv: hep-th/0611221v1

Bernard, D., & Folacci, A., Phys. Rev. D, 34, 2286 (1986)

Birrel, N. D., & Davies, P. C. W., Quantum Fields in Curved Space (Cambridge University Press, Cambridge, 1982)

R. Bousso, R. & Freivogel, B. (2006), "A paradox in the global description of the multiverse," arXiv:hep-th/0610132v2

Bunch, T. S., and Davies, P. C. W., Proc. R. Soc. London A, 360, 117 (1978)

Burgess, Gelett, "The Purple Cow" (1895)

Carlip, S., "Transient Observers and Variable Constants or Repelling the Invasion of the

Boltzmann Brains," arXiv: hep-th/0703115v5 (2007)

Dyson, L., Kleban, & Suskind, L., "Disturbing implications of a cosmological constant," JHEP 10, 011, arXiv:hep-th/0208013 (2002)

Farhi, E., Guth, A. H., & Guven, J., Nucl. Phys. B339, 417 (1990)

Garriga, J. & Vilenkin, A., "Recycling universe," Phys. Rev. D. 57, 2230 (1998), arXiv:astro-ph/9707292

Gibbons, G. W., & Hawking, S. W., Phys. Lett. 78B, 430 (1978)

Gott, J. R. Nature, "Creating Open Universes from de Sitter Space," 295, 304 (1982)

Gott, J. R., Nature, "Implications of the Copernican Principle for our future Prospects," 363, 315 (1993)

Gott, J. R., *Time Travel in Einstein's Universe*, Houghton Mifflin, Boston (2001)

Gott, J. R., & Li, L-X., Phys. Rev. D., 58, 3501 (1998), astro-ph/9712344

Gott, J. R., & Statler, T. S. Physics Letters, 136B, 157 (1984)

Guth, A. H., & Weinberg, E. J., Nucl. Phys. Rev. D 26, 2681 (1982)

Hartle, J. B., & Srednicki, M., "Are We Typical?" arXiv:0704.2630v3 [hep-th], (2007)

Linde, A. D., Phys. Lett. 129B, 177 (1983)

Linde, A., "Sinks in the Landscape, Boltzmann Brains, and the Cosmological Constant Problem," (2007) arXiv:hep-th/0611043v3

Narnhofer, H., Peter, I., & Thirring, W., Int. J. Mod. Phys. B 10, 1507 (1996)

Page, D., "Thermal stress tensors in static Einstein spaces," Phys. Rev. D., 25, 1499 (1982)

Page, D, "Is our universe likely to decay within 20 billion years?" (2006) arXiv:hep-th/0610079

Rindler, W., Am. J. Phys., 34, 1174 (1966)

Spergel, et al., ApJS, 170, 377 (2007) arXiv: astro-ph/06034449

Steinhardt, P. J., & Turok, N, "A Cyclic Model of the Universe," (2002) arXiv:hep-th/0111030v2

Unruh, W. G., Phys. Rev. D., 14, 870 (1976)

Vilenkin, A., "Making predictions in eternally inflating universe," Phys. Rev. D. 52, 3365, arXiv:gr-qc/9505031 (1995)

Vilenkin, A. "Freak observers and the measure of the multiverse," Journal of High Energy Physics, DOI: 10.1088/1126-6708/2007/01/092, arXiv:hep-th/0611271v2 (2006)